\documentclass[]{spie}  

\usepackage{amsmath,amsfonts,amssymb}
\usepackage{graphicx}
\usepackage[colorlinks=true, allcolors=blue]{hyperref}
\usepackage{siunitx}
\usepackage{xcolor}
\usepackage{float}
\usepackage{aas_macros}
\usepackage{svg}
\graphicspath{{figures/}{./}}

\title{Cherenkov Photon Background for Low-Noise Silicon Detectors in Space
}

\author[a, b, f]{Manuel E. Gaido}
\author[b, f]{Javier Tiffenberg}
\author[b, c]{Alex Drlica-Wagner}
\author[b]{Guillermo Fernandez-Moroni}
\author[d]{Bernard J. Rauscher}
\author[e]{Fernando Chierche}
\author[a, f]{Darío Rodrigues}
\author[a, f]{Lucas Giardino}
\author[b]{Juan Estrada}
\affil[a]{Universidad de Buenos Aires, Buenos Aires, Argentina}
\affil[b]{Fermi National Accelerator Laboratory, Batavia IL, United States}
\affil[c]{University of Chicago, Chicago IL, United States}
\affil[d]{NASA Goddard Space Flight Center, Greenbelt MD, United States}
\affil[e]{Universidad Nacional del Sur, Bahía Blanca, Argentina}
\affil[f]{LAMBDA, FCEyN, Argentina}

\authorinfo{Contact information:\\ Manuel E. Gaido: \href{manu.gaido@gmail.com}{manu.gaido@gmail.com}, \\  Javier Tiffenberg: \href{javiert@fnal.gov}{javiert@fnal.gov}, \\ Alex Drlica-Wagner: \href{kadrlica@uchicago.edu}{kadrlica@uchicago.edu}}

\begin{document} 
\maketitle

\begin{abstract}
Future space observatories dedicated to direct imaging and spectroscopy of extra-solar planets will require ultra-low-noise detectors that are sensitive over a broad range of wavelengths. Silicon charge-coupled devices (CCDs), such as EMCCDs, Skipper CCDs, and Multi-Amplifier Sensing CCDs, have demonstrated the ability to detect and measure single photons from ultra-violet to near-infrared wavelengths, making them candidate technologies for this application. In this context, we study a relatively unexplored source of low-energy background coming from Cherenkov radiation produced by energetic charged particles traversing a silicon detector. In the intense radiation environment of space, energetic cosmic rays produce high-energy tracks and more extended halos of low-energy Cherenkov photons, which are detectable with ultra-low-noise detectors. We present a model of this effect that is calibrated to laboratory data, and we use this model to characterize the residual background rate for ultra-low noise silicon detectors in space. We find that the rate of cosmic-ray-induced Cherenkov photon production is comparable to other detector and astrophysical backgrounds that have previously been considered.
\end{abstract}


\section{INTRODUCTION}
\label{sec:intro}  

The search for potentially habitable Earth-like planets around Sun-like stars is one of the priority areas identified by the recent Decadal Survey on Astronomy and Astrophysics \cite{Astro2020}. The direct imaging and spectroscopy of such planets is the primary objective of NASA's Habitable Worlds Observatory mission concept \cite{Clery:2023}.
Such observations seek to detect ``biosignatures'', aspects of a planet's atmosphere or surface that may indicate presence of life \cite{Seager:2013}. However, direct spectroscopy of the atmospheres of Earth-like exoplanets presents an enormous technological challenge \cite{Crill:2017,Crill:2022}. The light of the host star greatly outshines the reflected light of the planet, which may have an average rate of several photons per hour \cite{Seager:2010,Rausher:2019}. In this regime of signal-starved observations and long integration times, ultra-low-noise detectors are vital. In particular, recent technology reports have stated a need for detectors which can operate in visible to near-infrared (NIR) wavelengths with a readout noise of ${<}\,0.1\,\text{e}^- \text{rms}/\text{pix}$ \cite{Crill:2017,Crill:2022}. Given the precision required for these observations the limited population of potentially observable exo-Earths \cite{Astro2020}, it is critical to identify and characterize sources of detector, instrumental, environmental, and astrophysical backgrounds in order to accurately estimate the signal-to-noise ratio (SNR) of future exo-planet observations. 

Previous studies have considered several backgrounds that will affect the ability of a future space telescope to directly image and perform spectroscopy of exo-planets.
These include detector noise sources including readout noise, dark current, and clock induced charge \cite{Shaklan:2013, Crill:2017, Rausher:2019, Crill:2022}.
On the other hand, instrumental and astrophysical backgrounds include residual speckle light, zodiacal light, and exo-zodiacal light \cite{Lacy:2019}.
Energetic cosmic rays are a familiar and well-characterized environmental background when operating in the intense radiation environment of space. 
The primary cosmic ray track can be relatively easily identified through high charge occupancy, which allows them to be masked and effectively removed from science images. 
While the cosmic ray rate sets important limitations on instrument design and operation (i.e., radiation hardness, maximum exposure time, effective masked area per exposure), it is generally assumed to have a negligible impact on the single-photon background rate.
Here, we identify and characterize Cherenkov photon emission from energetic cosmic rays as a previously unstudied source of background for the detection and measurement of exo-Earths at visible/near-IR wavelengths.
Cherenkov photons emitted at optical and near-infrared wavelengths can travel a significant distance in silicon detectors before depositing their energy in pixels that are outside conventional cosmic-ray masks. 
We find that for conventional cosmic-ray masking approaches, the Chernekov photon background is comparable in magnitude to other astrophysical and detector backgrounds that are currently considered.

\subsection{CHERENKOV RADIATION}

\begin{figure} [t!]
\begin{center}
\begin{tabular}{c} 
\includegraphics[width=0.5\textwidth]{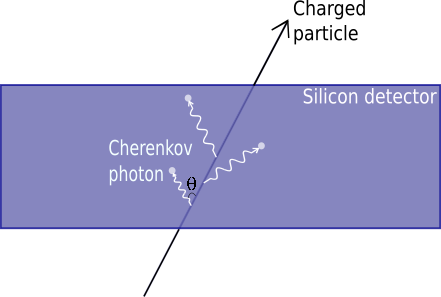}
\hspace{2cm}
\includegraphics[width=0.3\textwidth]{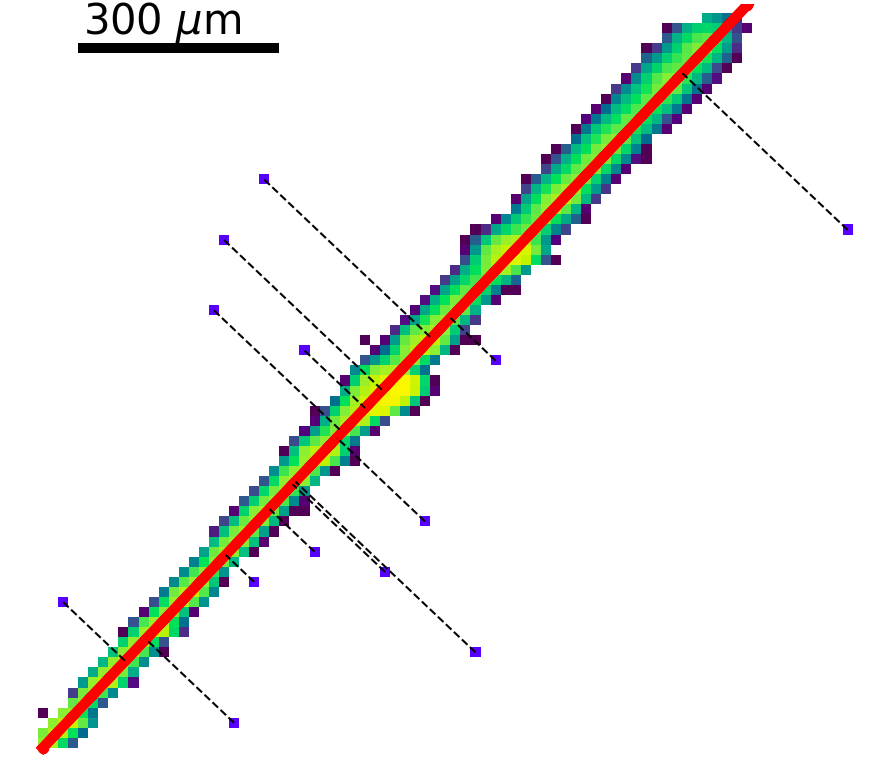}
\end{tabular}
\end{center}
\caption{(Left) Schematic representation of Cherenkov emission in silicon. (Right) Measured muon track with best fit track line (red line) and distances to Cherenkov photons measured orthogonal to the track (black lines).}
\label{fig:schematic-cherenkov}
\end{figure}

The Cherenkov effect \cite{Cherenkov:1934, Cherenkov:1937} is the process by which a charged particle crossing a polarizable material medium emits electromagnetic radiation (Fig. \ref{fig:schematic-cherenkov}). In order for Cherenkov emission to occur, the charged particle must be traveling faster than the speed of light in the material. The electric dipoles of the material's molecules are excited, emitting a constructively interfering electromagnetic conical wavefront. This effect is analogous to a sonic boom \cite{Jelley:1955}. 

The Frank-Tamm formula\cite{Frank:1937} provides the spectrum of emitted electromagnetic energy as a function of particle's charge, $q$, velocity, $\beta < 1$, and the dialectric function of the medium, $\epsilon \left( \omega \right)$, dependent on the emitted photon frequency, $\omega$. The differential number of photons produced per unit length per unit frequency is described as,\cite{Budini:1953}
\begin{equation}\label{eq:cherenkov}
    \frac{d^2 N_{\gamma}}{dx d \omega} = q^2 \alpha \left(  1 - \frac{\text{Re} \{ \epsilon \left( \omega \right) \} }{\beta^2 | \epsilon \left( \omega \right) |^2} \right), \qquad \beta > \frac{1}{\sqrt{\text{Re} \{ \epsilon \left( \omega \right) \} }},
\end{equation}
\noindent where $\alpha$ is the fine-structure constant.
Cherenkov radiation is emitted in a conical wavefront with it's vertex at the charged particle's current position and a characteristic Cherenkov angle, $\theta_{c}$, measured between the emitted photon direction and the source particle's velocity, given by
\begin{equation}\label{eq:theta-ch}
    \cos (\theta_{Ch}) = \frac{\sqrt{\text{Re}\{ \epsilon \left( \omega \right)\}}}{\beta \vert \epsilon \left(  \omega \right)\vert}.
\end{equation}
\noindent The dielectric function depends on the material's characteristics, and in silicon there is a sharp cutoff in Cherenkov photon energies at $\sim 4 \si{\electronvolt}$ \cite{Du:2022}. After Cherenkov photons are emitted, the distance that they travel from the charged track before being absorbed is set by the photon attenuation length,\cite{Du:2022}
\begin{equation}\label{eq:attenuation}
\ell_\gamma = \frac{1}{\omega \sqrt{ 2 |\epsilon(\omega)| - 2 \text{Re}\{ \epsilon(\omega)}\}}.
\end{equation}
Depending on their energy (i.e., the frequency, $\omega$), Cherenkov photons can be promptly absorbed or travel a considerable distance from the charged particle track.  Photons with energies $\gtrsim 2 \si{\electronvolt}$ have an absorption length that is smaller than the usual track width ($\lesssim 15\,\mu$m $ = 1\,\text{pix}$), making them invisible in an image taken by a silicon detector such as a CCD. On the other hand, photons with energies of $\lessapprox 1.2 \si{\electronvolt}$ have an absorption length that is significantly greater than typical detector thickness ($\gtrsim 100\,\mu$m) and commonly escape the detector without interacting.
In the intermediate energy range between $1.2\si{\electronvolt} \lesssim E \lesssim 2\si{\electronvolt}$, Cherenkov photons will be absorbed by the silicon generating an electron-hole pair via the photo-electric effect, which we refer to as a ``single electron event'' (SEE) \cite{Ramanathan_2020}. The production of SEE from Cherenkov photons has recently been studied as an important background in dark matter direct detection experiments using CCDs.\cite{Du:2022}

\section{MODEL AND VALIDATION}

To model Cherenkov emission in silicon, we first integrate Eq.~\ref{eq:cherenkov} in frequency space by setting a constant value of particle vector velocity $\boldmath{\beta}$. The range of integration is determined by the photon frequencies that satisfy the Cherenkov emission condition for a given velocity (Eq.~\ref{eq:cherenkov}). Due to the nature of silicon's dielectric functions \cite{Du:2022, GREEN20081305}, integration always takes place in a single, continuous frequency interval

\begin{equation}
    N_{\gamma} = \int_{\text{track}}\int^{\omega_{max}}_{\omega_{min}} \frac{d^2 N_{\gamma}}{dx d \omega} d \omega dx.
\end{equation}

\noindent We use this as the total emission per trajectory with constant values (i.e. velocity and particle type) by drawing from a Poisson distribution. Next, we assign each of these photons an initial position, direction and energy. The initial position is uniformly distributed along the particle's given trajectory and direction is set by Cherenkov's emission angle, $\theta_{Ch}$ (Eq.~\ref{eq:theta-ch}), and a random rotation angle, $\phi$, around the particle's direction. Photon energy is drawn from a uniform distribution ranging over the allowed emission frequencies. This approximation is accurate for ultra-relativistic particles that make up the vast majority of particles responsible of Cherenkov emission. Emitted photons will propagate through the material a random distance in their assigned direction, determined by Eq.~\ref{eq:attenuation}. Moreover, we assign emitted photons a probability of $0.25$ of being absorbed (generating an electron-hole pair) when they reach a surface of the detector, despite not having reached their pre-determined propagation distance. We plan to implement a detailed modeling describing reflection and transmission in a future work. 

Our Cherenkov model works in combination with the GEANT4 software\cite{GEANT4:2002}, which is used to perform a detailed propagation of charged particles in matter. By defining a silicon detector geometry in GEANT4, we are able to simulate events within the detector's sensitive volume. Events begin with a primary particle generated from a user-defined spatial and angular distribution and spectrum. From this, the primary particle interacts with the material and generates secondary particles, which are also tracked. For each particle, we store each step of its propagation through the silicon detector in an output ROOT file. A step contains initial and final position, velocity, deposited energy and particle type of each component that makes up the complete passage of the primary particle and any secondaries through the detector. The charge deposited by each step is then diffused in the image, simulating the drift of charge carriers generated in the bulk of the detector, with it's spread being positively correlated with the depth of interaction \cite{PhysRevLett.125.171802}. With this information, we are able to simulate charged particle events and append their corresponding Cherenkov emission using the modeling framework described above. We create a simulated image for the defined detector geometry containing both the charged particle track and corresponding low-energy halo of Cherenkov emission. 

\begin{figure}[t!]
\begin{center}
\begin{tabular}{c} 
\includegraphics[scale=0.45]{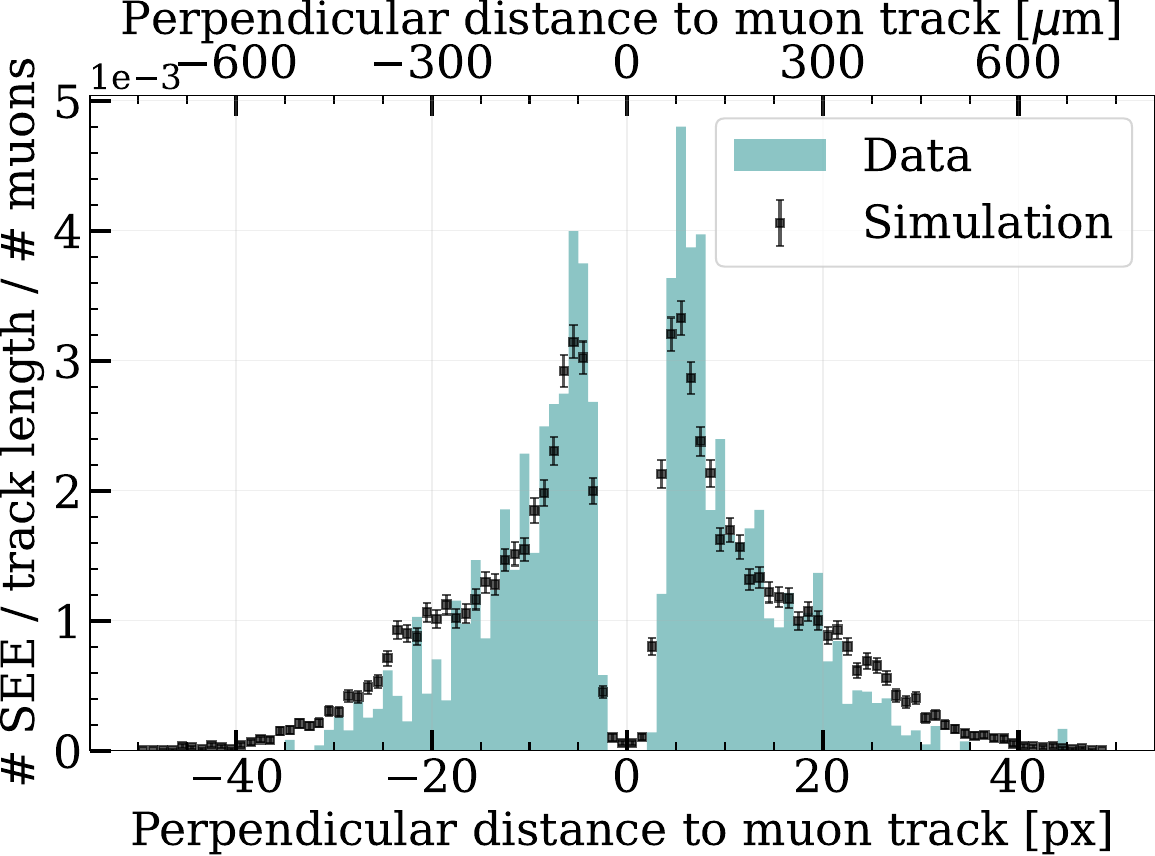}
\end{tabular}
\end{center}
\caption{Distribution of perpendicular distances between muon tracks and SEE for images taken by Giardino et al. (2022) \cite{Giardino:2022} (teal histogram) and simulations from our model (black points). Simulations include the additional dark current rate that was estimated from the original dataset in order to keep conditions as similar as possible.}
\label{fig:see-distance-distribution}
\end{figure}   

We validate our model against real cosmic ray data collected with an ultra-low-noise silicon detector capable of counting SEE associated with Cherenkov emission. In particular, Giardino et al. (2022)\cite{Giardino:2022} present a dataset of cosmic ray muons observed at surface level with an ultra-low-noise Skipper CCD (6144$\times$866, 15 $\si{\micro\meter}$ pixels and 675 $\si{\micro\meter}$ thickness). The Skipper-CCD was designed at LBNL and fabricated at Teledyne DALSA Semiconductor on high-resistivity silicon. In order to see the low-energy halo around high-energy events, the Skipper CCD was operated in ``Smart Skipper'' mode with both energy and region of interest readout \cite{Chierchie:2021}. This readout configuration provides single-photon/single-electron resolution around high energy events while minimizing the readout time and dark current. Filtering for atmospheric muons, we can re-create these events in GEANT4 by setting their incidence direction as measured in the original images and assuming that their energy is the average atmospheric muon energy ($\sim 4 \si{\giga\electronvolt}$) \cite{Autran:2018}. By simulating several realizations of the same accepted muon events, we are able to compare data and simulations by the distribution of SEE as a function of perpendicular distance to its source particle (Fig.~\ref{fig:see-distance-distribution}). We measured the dark current event rate and included this effect in our simulation. 

We find good agreement between data and simulations. We now extend our validated model to predict Cherenkov emission from cosmic ray protons for detectors in space. 

\section{SIMULATION OF SPACE ENVIRONMENT}

\begin{figure} [ht]
\begin{center}
\begin{tabular}{c} 
\includegraphics[scale=.4]{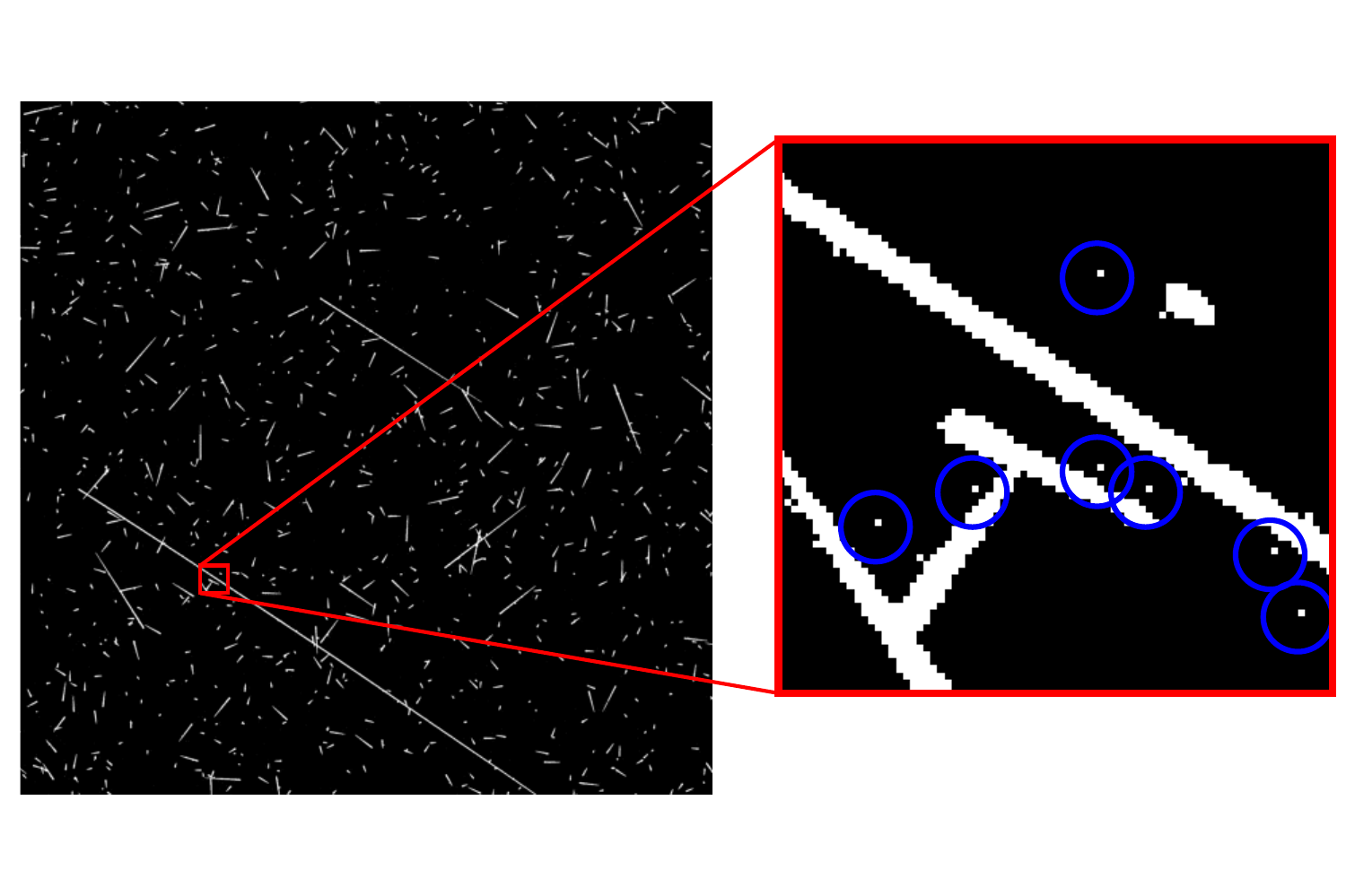}
\end{tabular}
\end{center}
\caption{Simulated $30 \si{\second}$ exposure for a 2k$\times$2k, $15 \si{\micro\meter}$ pixel, $250 \si{\micro\meter}$ thick silicon detector in LEO including cosmic-ray proton tracks and secondary Cherenkov photons. The expected proton rate is $\sim 950$ particles per image, each of which has some probability of generating one or more Cherenkov photons (inset; blue circles).
}
\label{fig:example-exposure}
\end{figure}

The cosmic-ray environment in low Earth orbit (LEO) and at the L2 Lagrange point consists primarily of protons, with steeply falling energy spectrum~\cite{10.1093/ptep/ptaa104, hillas2006cosmic}. We simulated the cosmic-ray environment in LEO for a red-sensitive silicon detector with a thickness of $250\,\si{\micro\meter}$ comprised of 2k$\times$2k, $15 \si{\micro\meter}$ pixels. This detector thickness is characteristic of a Skipper CCD that provides high quantum efficiency at near-infrared wavelengths \cite{Drlica-Wagner:2020, Marrufo:2024}. We simulate particles according to an isotropic angular distribution and uniform spatial distribution. In order to simulate a realistic data taking strategy, we break up a 1 hour observation into $120 \times 30 \si{\second}$ exposures. Based on the average cosmic-ray rate in LEO, a typical $30 \si{\second}$ exposure for this detector geometry will contain ${\sim} 1000$ high-energy events, $\gtrapprox 95 \%$ of which are protons~\cite{10.1093/ptep/ptaa104}. We simulate a dataset of over $2 \times 10^5$ unique protons, from which we sample $120$ unique images each with a $30 \si{\second}$ exposure time (Fig.~\ref{fig:example-exposure}). Alpha particles are the second most abundant species, but we restrain ourselves only to protons for simplicity.

\begin{figure}[H]
\begin{center}
\includegraphics[width=0.9\textwidth, trim={0cm 9cm 0cm 0cm}, clip]{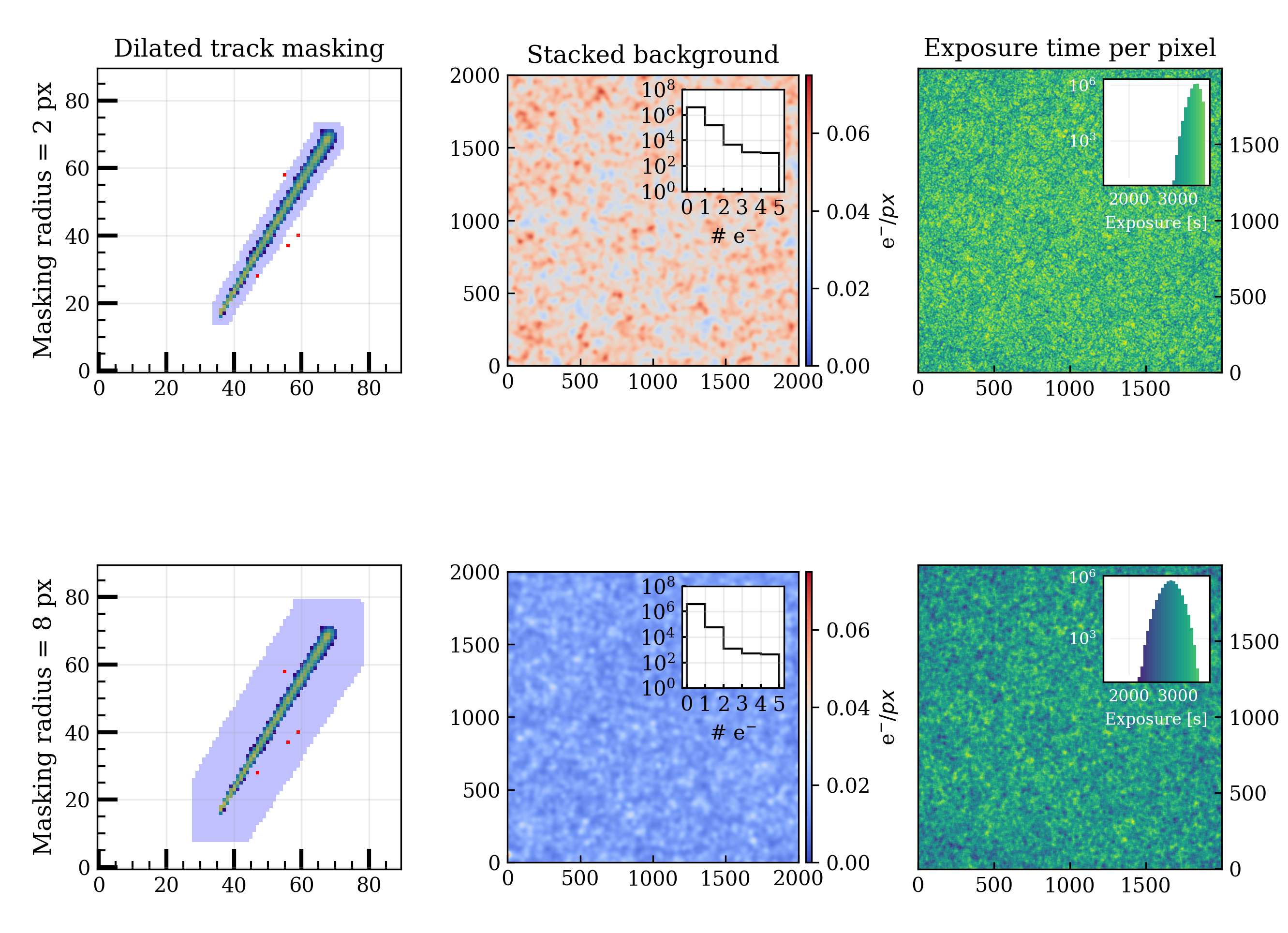}
\includegraphics[width=0.9\textwidth, trim={0cm 0cm 0cm 9.5cm}, clip]{comparative-panel-2x2bin_blurr.png}
\end{center}
\caption{(Left) Examples of small (2 pixel; top) and large (8 pixel; bottom) masking radii around a simulated proton track and secondary Cherenkov photons. (Middle) Simulated residual Cherenkov photon rate after masking and combining 120$\times$30\,s exposures, binned in 2$\times$2 blocks and convolved with a Gaussian kernel. (Right) Effective exposure time for the combination of 120$\times$30\,s exposures after masking. 
}
\label{fig:comparative-panel}
\end{figure}

\begin{figure}[t!]
\begin{center}
\begin{tabular}{c} 
\includegraphics[scale=.45]{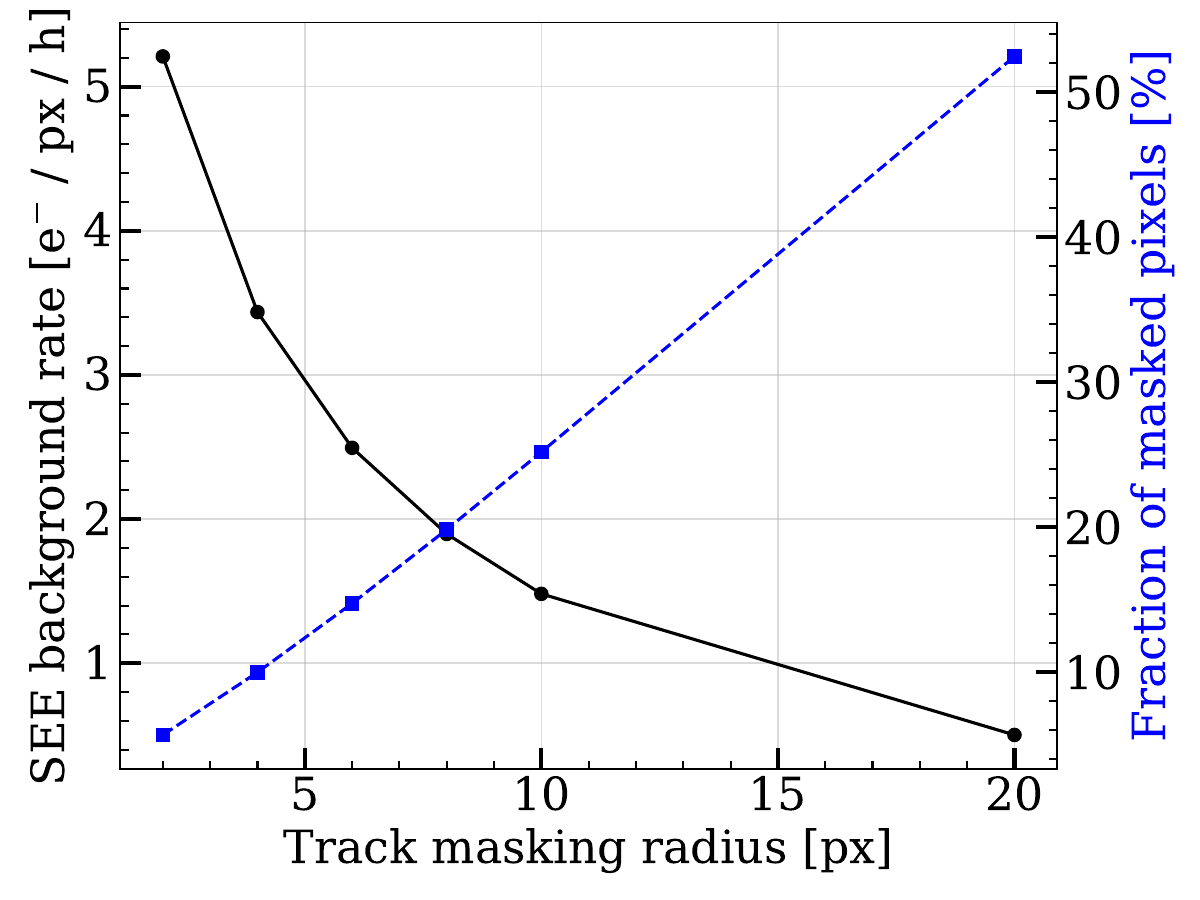}
\end{tabular}
\end{center}
\caption{Residual Cherenkov background rate (solid black curve) and fraction of masked pixels (dashed blue curve) as a function of the masking radius for a 2k$\times$2k detector. A higher fraction of masked pixels translates into a lower effective exposure time.
}
\label{fig:rate-exposure}
\end{figure}

For processing, tracks are identified in each exposure by setting an energy threshold and masking pixels with values that are above this threshold. This mask is then dilated by a fixed number of pixels outward (i.e., the ``masking radius''). All of the exposures are stacked after removing high-energy tracks to obtain the combined observation. 
Pixels in the combined observation will have different effective exposure times due to masking, and Cherekov photons that reside outside the track masks will contribute to the residual Cherenkov background rate.
For a single image, the residual unmasked Cherenkov photons and masked exposure time trace the cosmic-ray tracks. 
However, as many images are combined these distributions become more uniform due to the uniform, isotropic distribution of simulated cosmic rays.
The middle frame of Fig.~\ref{fig:comparative-panel} gives a qualitative picture of the spatial distribution of residual Cherenkov events in the combined image after masking regions around high-energy events. The small masking radius produces a higher mean value and corresponding shot-noise fluctuations in the image, while the larger masking radius results in a lower Cherenkov background rate with smaller variations. We scan over different masking radii and show the trade off between the fraction of masked pixels against the Cherenkov background rate in Fig.~\ref{fig:rate-exposure}.

\section{RESULTS AND CONCLUSIONS}

Based on our model of the cosmic-ray environment in LEO, we find that the rate of SEE generated by Cherenkov photons is  $\gtrsim 1 \text{e}^{-}/\text{pix}/\si{\hour}$ assuming cosmic-ray-track  masking radii ranging from 2 to 20 pixels (Fig.~\ref{fig:rate-exposure}). 
We note that the upper bound of ${\sim} 20$ pixels is much larger than conventional cosmic-ray masking algorithms \cite{vanDokkum:2001} and leads to a significant (${\sim} 50\%$) reduction in total signal integration time. 
Furthermore, when searching for very faint signals ($\ll 1$ photon per image), alternative approaches to removing particle background such as median filtering or difference imaging do not work.

\begin{figure}[t!]
\begin{center}
\begin{tabular}{c} 
\includegraphics[scale=.45]{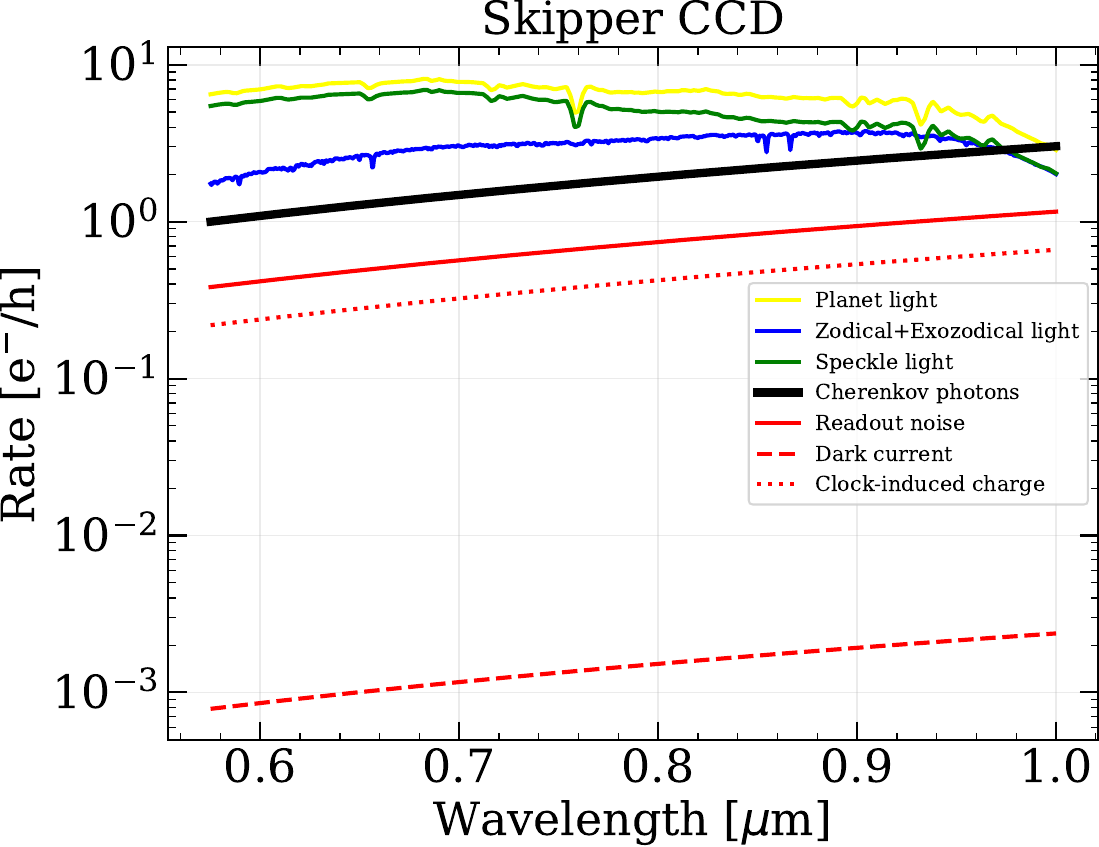}
\end{tabular}
\end{center}
\caption{Signal and background sources for direct exo-planet spectroscopy using a Skipper CCD as a function of wavelength. This figure was generated to mimic observations from the Roman-CGI in IFS mode observing a Jupiter-radius planet orbiting a 5.0\,mag G0V star at a separation of 3.8\,AU observed at a distance of 10\,pc using software provided by Lacy et al. (2019)\cite{Lacy:2019}. Skipper CCD detector characteristics were taken from the literature.\cite{Drlica-Wagner:2020, Barak:2022}. The residual Cherenkov rate used here is $2.5\,e^- / {\rm pix} / {\rm h}$, correspondent to a $10$\,pix masking radius.
}
\label{fig:signal-background-skipper}
\end{figure}

We compare the rate of Cherenkov photons produced in a thick Skipper CCD to the expected exoplanet signal and other sources of background.
To estimate the various signal and background contributions, we follow the model described in Lacy et al. (2019)\cite{Lacy:2019}, which was developed to simulate the characterization of exoplanet atmospheres with the Coronagraphic Instrument (CGI) on the {\it Nancy Grace Roman Space Telescope}.
Lacy et al.\ provide a detailed description and open source code\footnote{\url{https://github.com/blacy/direct-imaging-sims}} for estimating signal and background sources for exoplanet spectroscopy, with a focus on the giant exoplanets that will be observable by the {\it Roman} CGI.
Background sources include detector backgrounds, such as readout noise, clock induced charge, and dark current (intrinsic sources), as well as speckle and zodiacal light (extrinsic sources).
Lacy et al.\ focus on detector backgrounds associated with the Teledyne-e2v CCD201 EMCCD detectors that were baselined for the {\it Roman}-CGI at the time of their study.
We modify their code to instead use detector parameters for a thick, fully depleted Skipper CCD, taking literature values for the Skipper CCD quantum efficiency, dark current, and spurious clock induced charge \cite{Drlica-Wagner:2020, Barak:2022}.
In Fig.~\ref{fig:signal-background-skipper} we present the resulting estimates for the signal and background rates for the {\it Roman}-CGI operating in integral field spectrograph (IFS) mode for an observation of a Jupiter-radius planet orbiting a 5.0\,mag G0V star at a separation of 3.8\,AU observed at a distance of 10\,pc.
It can be seen that the Cherenkov photon rate dominates over other Skipper CCD detector background contributions and is comparable to astrophysical backgrounds. The masking radius assumed for the rate presented in Fig.~\ref{fig:signal-background-skipper} is of $10$\,pix.

We emphasize that the Cherenkov radiation is an irreducible source of background for \emph{all} silicon detectors, as well as any other detectors using dielectric materials where Cherenkov radiation can be emitted.  
For silicon detectors with coarse timing resolution, including EMCCDs, Skipper CCDs, and MAS CCDs, Cherenkov photons in the energy range of $1.2$\,eV to ${\sim} 2$\,eV are indistinguishable from astronomical source photons, and can only be identified probabilistically due to their higher occurrence rate close to cosmic-ray particle tracks.
The cosmic-ray track length, and thus the Cherenkov photon rate, can be reduced with thinner detectors; however, thin detectors will be less sensitive to near-infrared photons, which is a critical regime for identifying bio-signatures in exoplanet atmospheres.\cite{Rauscher:2022}


\appendix    

\acknowledgments 

This work was partially funded by Fermilab LDRD (2019.011, 2022.054), NASA APRA (No. 80NSSC22K1411), and the Heising-Simons Foundation (\#2023-4611).
This manuscript has been authored by Fermi Research Alliance, LLC under Contract No. DE-AC02-07CH11359 with the U.S. Department of Energy, Office of Science, Office of High Energy Physics. 
The CCD development work was supported in part by the Director, Office of Science, of the DOE under No.~DE-AC02-05CH11231.
The United States Government retains and the publisher, by accepting the article for publication, acknowledges that the United States Government retains a non-exclusive, paid-up, irrevocable, world-wide license to publish or reproduce the published form of this manuscript, or allow others to do so, for United States Government purposes.

\bibliography{report} 
\bibliographystyle{spiebib} 

\end{document}